# User Profiling Trends, Techniques and Applications

Sumitkumar Kanoje, Sheetal Girase, Debajyoti Mukhopadhyay
Dept. of Information Techonlogy, Maharashtra Institute of Technology, Pune
sumitkanoje@gmail.com, sheetal.girase@mitpune.edu.in, debajyoti.mukhopadhyay@gmail.com

**A B S T R A C T**

**The Personalization of information has taken recommender systems at a very high level. With personalization these systems can generate user specific recommendations accurately and efficiently. User profiling helps personalization, where information retrieval is done to personalize a scenario which maintains a separate user profile for individual user. The main objective of this paper is to explore this field of personalization in context of user profiling, to help researchers make aware of the user profiling. Various trends, techniques and Applications have been discussed in paper which will fulfill this motto.**

## I. INTRODUCTION

Personalization has got a lot importance in the field of computer science, being specific in the applications of Recommendation Systems. There are many users which a recommender system needs to deal with, and each user has their own preference requirements. Recommender System needs to fulfill the requirements of every user by recommending user specific items to them or by modifying itself as per the needs of the user. So user profiling helps recommender system to know user requirements and behave in accordance of them.

User Profiling can be defined as the process of identifying the data about a user interest domain. This information can be used by the system to understand more about user and this knowledge can be further used for enhancing the retrieval for providing satisfaction to the user. User profiling has two important aspects as efficiently knowing user and based on those recommending items of his interest. This paper tries to explore all the aspects of a User Profiling system in the Recommender System.

This paper surveys user profiling in three main contexts. First the trends in the user profiling are identified to show how the user profiling evolved from the recommender systems, then we go on discussing the techniques used for profiling the users, finally the paper gives some case studies on how User Profiling has been applied in various domains.

## II. TRENDS

Capturing information about users and their interest is the main function of user profiling. Much of the research has been done on profiling in the field of recommender system and various profiling techniques have been evolved around the time.

User profiling in the general evolved with the data mining & machine learning approach. Its root can be found in the knowledge data discovery model [1], many of the steps in KDD model resemble to the steps involved in the user profiling process. This paper calls user profiling as User Data Discovery (UDD) model.

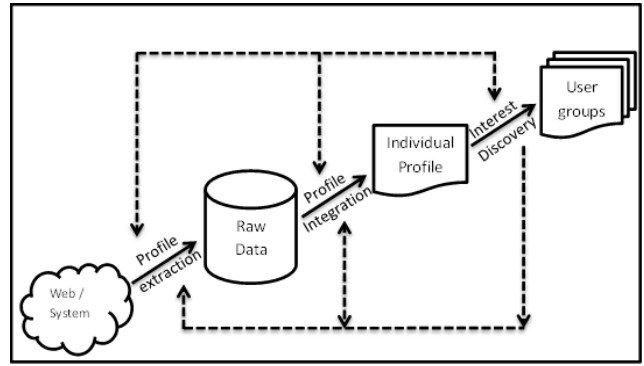

**Fig 1.** User Data Discovery

To show the resemblance between KDD process and UDD Process, a modified KDD model is described in the figure.

In the knowledge Data Discovery model there is already a lump sum of data available to the system whereas in UDD model the system has very few information about the user and the system has to timely gain the knowledge about a user which could be used for the further operations.

Gaining knowledge is the main aim of both the systems but the way they perform this task is the matter of point here. One which needs to extract knowledge from a big chunk of data i.e. it has to process the available data to find out something which is interesting whereas the other case includes the data which has very less amount of data, and still it has a crucial job of behaving as per the user requirements.

User profiling in general started with the just information retrieval and collection of the user's information. Older systems were more concerned about getting data directly from the users that is the system was explicitly asking the users about the data which is needed. But this method is not considered efficient as the user is never interested in directly giving the input so now day's research is more focused on profiling user's data implicitly based on some actions performed by the user, it may also be referred as behavioral user profiling. There have been many research happened on this and we can identify three main approaches of user profiling as follows

A. **Explicit User Profiling**

Danny Poo [3] explains the static profiling as a process of analyzing users static & predictable characteristics. In this approach users behavior is predicted by analyzing the user's available data. This data comes usually by filing the online forms or from surveys etc. This is also referred to as static profiling or factual profiling. There are some problems when we only depend on the use of implicit profiling as users are not interested to reveal their information to anyone as they are concerned about their privacy or the form filling process might be tedious which the user tries to avoid it. Hence the accuracy of using this type of profiling degrades according to the time period.

B. **Implicit User Profiling:**

Danny Poo [3] also explains dynamic profiling approach in which Instead of concentrating on the current information we have about the user this approach relies more on what we have known about user in future i.e. systems tries to learn more about the user. So such type of system is also called as Behavioral profiling, Adaptive Profiling and now days it's more referred to as Ontological Profiling of the user. Different types of filtering techniques are also used in such profiling. A lot of research literature can be found which discussed some filtering techniques, some of which could be Rule based filtering, Collaborative filtering and Content based filtering techniques [16].

C. **Hybrid User Profiling:**

This type of user profiling combines the advantages of both Implicit as well as explicit user profiling. I.e. it takes into consideration both the static characteristics about a user and also goes retrieving the behavioral information regarding a user. This approach helps profiling more efficient and maintains the accuracy of temporal information as information gets updated temporally.

## III. TECHNIQUES

By looking at different research literature we can be inferred that a user profiling system can be subdivided into subtasks like profile extraction, profile integration and user interest discovery. Now we will look for different techniques used for each of these tasks.

A. **Profile extraction**

Profile extraction is nothing extracting the useful information about a user from different sources. For this cause many methodologies and models have been used by different researchers. Some of which can be mentioned as extracting data from web, extracting data from social media websites as well as some user behavior based techniques which help user profiling system gather interesting information about users.

Jie Tang [2] has explained the task of extracting profiles involves finding relevant pages from web & then finding relevancy using support vector machine algorithm. In this technique Jie tang [2] used preprocessing on the relevant information using linear conditional random fields and tree structured conditional random fields. It has been also found that the Tree Structured Random fields perform well as compared to linear random fields.

In another approach by Jiwei Li [9] has tried to extract user information from social networking websites like Twitter, Google plus and Facebook. In this they used present a weakly supervised approach to user profile extraction from Twitter [9]. Data from these websites are used as a distant source of supervision for extraction of their attributes from user-generated text. In addition to traditional linguistic features used in distant supervision for information extraction, this approach also takes into account network information, a unique opportunity offered by social media.

Yoshinori Hijikata [6] has done a very interesting study in which user's mouse operation are traced to extract text parts which the user might be interested in from the whole text of the Web page. An implicit relevance feedback [11] is given to the system by which system can generate relevant recommendations. Some of the main operations that are being pointed out specifically are

- Text tracing: Moving the mouse pointer along a sentence while reading.
- Link pointing: Positioning the mouse pointer on a link, but not clicking the link.
- Link clicking: Clicking on a link to move to another page.
- Text selection: Selecting text by dragging the mouse pointer.
- Scrolling: Scrolling a window at a certain speed.
- Bookmark registration: Registering a page as a bookmark.
- Saving: Saving an HTML document.
- Printing: Printing a page.
- Window movement: Moving a window of the Web browser.
- Window resizing: Changing the window size of the Web browser.

B. **Profile Integration:**

After the extraction of relevant information is done there might be sometimes a problem of data cleaning. Its might happen that some of the data that has been collected might be a duplicate data or it may look like duplicate data but that might be a unique data. So to resolve this issue unique and duplicate data must be identified which will help profiling system in the next subtask.

One of such problem is identified by Jie Tang et.al. [2] Which they referred it as Name Disambiguation Problem. A very efficient technique is being used to resolve the name disambiguation problem viz. Hidden Markov Model. Hidden Markov Random Field (HMRF) is a member of the family of MRF and its concept is derived from Hidden Markov Models (HMMs) [Ghahramani and Jordan 1997].

C. **Interest Discovery [10]:**

After getting some information has been collected about the users it is necessary to chop down the users into different classes, as this will help giving feedback to the system. This can be done by grouping the users based on their behaviors into some classes. This technique is also called as filtering and much of the research has been done on these techniques. The filtering techniques are discussed in more details below.

1. **Content Based filtering**

Content-based filtering also called as cognitive filtering. It recommends items based on a comparison between the content of the items with a user profile and selects those items whose content best matches with the content of another item. The content of each item is represented as a set of descriptors or terms, typically the words that might be associated with an item. The user profile is represented with the same terms and built up by analyzing the content of items which have been seen by the user. This technique mostly depends on the explicit ratings or preferences given by the users for particular item and tries to find users who have given similar ratings for the item; But in practical scenario users never tend to give explicit ratings to the system. Thus a mechanism is needed which will implicitly identify the rating or preference of a user for the particular item.

2. **Collaborative filtering**

This technique usually involves clustering the users with the similar interest groups. This is basically based on the idea that users who agreed for some items in the past are likely to agree for same in the near future too. So this technique organizes users with the similar interest into groups of peers thus enabling the above idea of recommending the articles within same group of users. The effectiveness of this technique is heavily dependent on the how well the clustering of profiles correlates the users.

Alternative approaches for filtering information have been proposed as well. Demographic filtering systems for example use demographic information such as location, age, gender and education to identify the types of users that like a certain item [8]. Also a similarity based fuzzy clustering for user profiling is also used by [5] which measures similarity between different users by analyzing web logs. In another approach [7] developed a questionnaire to assess aspects of perceptual preferences in regard to information processing, knowledge gain and learning, named Perceptual Preferences Questionnaire used user centric profiling based on perceptual preference questionnaire.

IV. **APPLICATIONS**

Recommender systems can be considered as the direct beneficiary of user profiling, user profiling is an important part of the recommender systems since earlier times. But nowadays user profiling is becoming common in many of the applications like Search Personalization [15], Adaptive Websites, Adaptive Web stores and Customer Relationship Management systems. Some of the case studies of such applications are given in this section.

User profiling for recommendation of research papers is an application where much of the work has been done. The system developed by Jae Tang viz. ArnetMiner [2] has divided the task of user profiling into three subtasks viz. profile extraction, integration and interest discovery. In a similar approach by Stuart Middleton [4] used one extra step viz. profile visualization to represent a profile generated by the system which used ontological approach.

e-Tourism based website [13] is another application which can be benefited by User Profiling. This system was able to deliver personalized information based on the location of the user. As the tourism business is totally dependent on the demographic information like location the system was able to provide recommendations of nearest tourist spots to a new user based on his location.

Energy management is a very important task of nowadays. There are many big enterprises which are facing the challenge of efficient and optimized energy management. The smart energy management system developed by [12] have proved to be efficient this task. Here they used user profiling and micro accounting for smart energy management.

Finding a job is one of the tedious jobs everyone has to do in his lifetime. So developing a system which will automatically recommend jobs to a user as per his qualification and experiences is one the great idea which this author has succeeded to implement. This system is called CASPER (Case-Based Profiling for Electronic Recruitment) [14]. The system takes into account users profiling information and recommends suitable jobs to every individual.

## V. CONCLUSION & FUTURE WORK

Personalization in context of User Profiling is not a new concept; much of the study that has been done in this area tries to better understand the user which will help making choice decisions easier by selecting an appropriate item in which the user is interested.

In the age of Big Data where information is growing in all respect, extracting knowledge out of it is a difficult task. According to Long-tail theory if a user has so many options to choose from, there is a danger of user getting lost in deluge of excess choices. This is where Social Information Discovery and User Profiling play an important role. Peoples tend to take their friends' opinion before making their own choice. So an implicit user profiling through social discovery will help resolving long-tail problem.